\documentclass[conference]{IEEEtran}
\usepackage[english]{babel}
\usepackage{amsmath,amssymb,amsthm}
\usepackage{graphicx,booktabs,array,multirow}
\usepackage[nosort,nocompress]{cite}
\usepackage[hidelinks]{hyperref}
\usepackage{url}

\newcommand{\E}{\mathbb E}
\newcommand{\Prb}{\mathbb P}
\newcommand{\inc}{\pi^{\mathrm I}}

\newtheorem{theorem}{Theorem}
\hypersetup{pdftitle={CARA: Exact Local Repair Queries for Cloud Consolidation},pdfauthor={Xiyang Zhang, Hongzhi Wang}}
\title{CARA: Exact Local Repair Queries\\for Cloud Consolidation}
\author{\IEEEauthorblockN{Xiyang Zhang\textsuperscript{1,2}, Hongzhi Wang\textsuperscript{1,*}}\IEEEauthorblockA{\textsuperscript{1}Harbin Institute of Technology \qquad \textsuperscript{2}Zhongguancun Academy\\s-zhangxiy24@bza.edu.cn, wangzh@hit.edu.cn\\\textsuperscript{*}Corresponding author}}
\begin{document}
\maketitle

\begin{abstract}
Cloud consolidation evaluates alternative workload placements to reduce overload under fixed host and migration budgets. A controller repeatedly queries the same demand history, but consumes only a small ranked result and one final repair. Shared aggregation and incremental assignment reuse individual computations but still reconstruct constraints and numerical certificates for many related queries. This preparation can erase the savings from skipping exact profiles. We present CARA, which compiles seed-universal movement constraints and family-wide load-error budgets, then acquires exact profiles according to certified ranking keys. The compiler leaves a migration-bounded assignment core and preserves ranked records, physical mappings and continuation state; a separate admission stage evaluates one frozen repair. On all 12,006 hourly series from four workload sources, CARA executes controlled queries 4.75 times faster than compact complete profiling and heterogeneous packed queries 3.05 times faster than family-local execution; across 300 budgeted consolidation decisions, it reduces excess over FFD by 35.8\% and attains the lowest excess among the compared policies.
\end{abstract}

\section{Introduction}
Cloud consolidation packs workloads onto fewer hosts, making the coincidence of their demands as important as their individual peaks. After deployment, a controller must repeatedly determine whether a small redistribution can reduce overload while retaining the same hosts and respecting a physical migration budget. Database consolidation and tenant-placement systems use workload histories to evaluate such decisions~\cite{curino2011kairos,schaffner2013rtp,taft2016step,cortez2017resourcecentral,poppe2020seagull,konig2022tenant,arora2023flexible,konig2023solver}. For a running placement, this evaluation lies on the response path: processing more useful alternatives within the available time can improve the repair that is eventually released.

The controller's inner loop is a workload-query problem. For each candidate, it joins workload identities with proposed host groups, aggregates their demands over time, and computes a reliability profile. It then matches those groups to physical hosts to determine the minimum migration count. Thousands of closely related candidates may feed a result pool containing only a few records, from which a separate selection stage chooses one action. Most complete profiles and provisional physical mappings are consequently never used. We study how to execute these repair queries with less computation and workload-data access while preserving the exact ranked result.

Multiquery optimization shares common expressions~\cite{roy2000multiquery,giannikis2012shareddb}; incremental views and compiled what-if evaluation reuse aggregates across related inputs~\cite{ahmad2012dbtoaster,budiu2023dbsp,kara2024fivm,mohammed2024fade}. Dynamic assignment updates an optimum after local cost changes, and matching reductions fix edges within an assignment~\cite{millstettey2007dynamic,korenwein2018reduction,morita2024update,grossmann2026hypergraph}. These techniques expose different units of reuse. Resident aggregates distinguish workloads by identity, whereas migration depends on their incumbent origins; constraints shared by successive repair families form a third level. Certified top-$k$ methods avoid scores that cannot affect the answer~\cite{fagin2001aggregation,sui2026cas}, but first need bounds on those scores. The unresolved cost lies in preparing movement constraints and numerical evidence repeatedly across these levels.

Throughout a decision, the incumbent is the deployed placement and fixed migration reference. A seed starts at the incumbent and may later be a retained candidate; it generates queries without moving workloads. Redistributing eight mutable workloads between two groups with fixed nonempty backgrounds yields 256 assignments. With four workloads originating on each of two incumbent hosts, these assignments have at most 25 origin-count vectors, hence 25 movement problems, but potentially 256 load profiles. Their useful reuse keys are therefore different. Local edits also need not imply a local assignment: an exchange can propagate through connected overlap support. Restricting matching to the two edited groups therefore requires a certificate fixing the remaining matches. Rebuilding that certificate for each pair loses the reuse offered by their common seed.

Avoiding full profiles introduces a second difficulty: the exclusion certificate must cost less than the profile it saves. Reliability combines threshold counts, changing active-host denominators and peaks, so a reusable load sum alone does not supply the ranking bound. Scanning unchanged hosts, summing each mutable subset and rebuilding rounding allowances can repeat much of the original work. Loose bounds retain unnecessary profiles; tighter bounds can become expensive to prepare. Reassociating sums can also change a capacity predicate or a ranking tie. Recent decision-oriented query systems direct expensive evaluations with cheaper evidence~\cite{lee2026evergreen,wang2026querycentric,hassanzadeh2026semviews,sui2026cas}. Repair queries need that evidence to share preparation across families while preserving native numerical results, physical ties and the duplicate state consumed by later queries.

\begin{figure*}[t]
\centering\includegraphics[width=\textwidth]{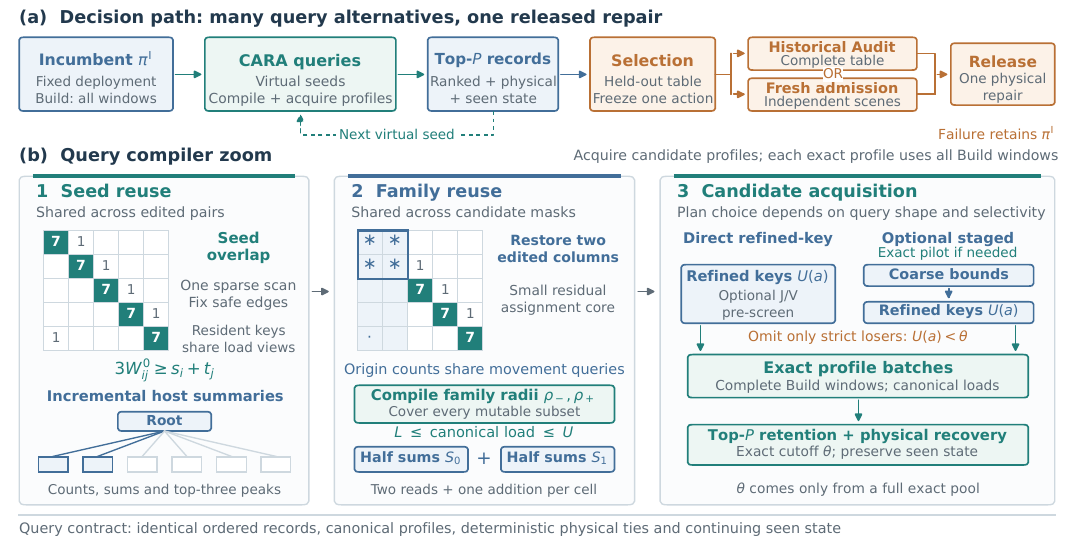}
\caption{CARA's decision path and query compiler. (a) Build queries update a ranked physical pool using virtual seeds; Selection freezes one action, followed by historical or fresh admission. (b) The compiler shares state at seed, family and candidate scopes. The illustrative overlap matrix has seven native and one cyclic resident per group; rows are incumbent hosts and columns are seed groups. Family budgets support both direct and optional staged acquisition. Only certified strict losers are omitted; exact profiles use every Build window. The cutoff comes from a full exact pool.}
\label{fig:method}
\end{figure*}

We present CARA, an exact local repair query processor with one-action certification. Figure~\ref{fig:method} connects its query engine to the complete repair decision. A seed compiles shared movement constraints and host summaries; a family specializes this state and compiles load-error budgets; ranking certificates then determine which candidate profiles to materialize. This data selection follows the ranked result: every acquired profile still uses the complete Build table. Selection freezes one retained physical repair, and admission evaluates its historical Audit table or independent fresh scenes before release.

Two compilation results make this flow economical. A seed-universal overlap certificate leaves an assignment core bounded by seed migration, even when the overlap support is connected. Family-wide numerical budgets enclose every candidate's prescribed stored loads using ordinary additions and subtractions, avoiding repeated error analysis. Direct refined-key and optional coarse-to-refined acquisition share these budgets and preserve the complete ranked physical answer and continuing seen relation. Once a repair is frozen, its migration components provide action-dependent support bounds for fresh-scene admission. The same resident dependencies thus guide both profile materialization during search and evidence acquisition for release.

Our evaluation includes every hourly series from four CloudCons sources, heterogeneous forecast-feasible packings at widths eight through sixteen, and complete budgeted consolidation decisions~\cite{zhang2026cloudcons}. All query comparisons charge compilation and preserve the same ranked records, physical outputs, and continuation state. CARA executes the controlled full-source queries 4.75 times faster than Compact and 1.55 times faster than sparse global assignment. Across 300 budgeted decisions, it reduces excess over FFD by 35.8\% and attains 4.9\% lower excess than a joint-metric annealer with 5.0\% less movement. The full-source budget study separately measures completed work, acquisition cost and overrun, connecting the query compiler to the controller's response budget.
\begin{itemize}
\item A shared seed-to-family compiler that separates resident and origin-count equivalence, with a seed-universal matching certificate and a migration-dependent bound on the residual assignment.
\item A family-scoped numerical compiler for exact profile certificates, backed by incremental host summaries and shared subset operands. Direct and staged acquisition plans preserve ranked and continuing query semantics.
\item Action-dependent admission bounds and a full-source evaluation connecting exact query efficiency, wider repair neighborhoods, and released repair quality.
\end{itemize}

\section{Repair Queries and Workload Data}\label{sec:problem}
We first specify what a repair may change and what a query must preserve. Workload and movement semantics determine eligibility, reliability profiles determine ranking, and the numerical contract fixes exact execution.

\subsection{Workload Data and Physical Movement}
A placement assigns $N$ workloads to $H$ nonempty homogeneous hosts of capacity $C$. We write $\inc$ for the running placement, which remains the movement reference throughout a decision. The demand relation $\mathcal D(i,b,t,x)$ records workload $i$ in demand window (scene) $b\in[B]$ at time $t\in[T]$; $\mathcal P_\pi(i,h)$ assigns it to a candidate host group. Joining these relations and aggregating by $(b,h,t)$ gives the load $\ell_{bht}(\pi)=\sum_{i:\pi(i)=h}x_{bit}$. A common forecast $\widehat x_{it}$ supplies the capacity predicate
\begin{equation}\label{eq:feasible}
\sum_{i:\pi(i)=h}\widehat x_{it}\le C+\tau\qquad\text{for every }h,t,
\end{equation}
where $\tau=10^{-12}$. Forecasting and uncertainty-aware packing can provide this input in several ways~\cite{calafiore2006scenario,nemirovski2007convex,campi2008exact,song2014chance,esfahani2018wasserstein,zhang2018ambiguous,cohen2019overcommitment,zhang2020branch,hyndman2021forecasting,borges2024scenarios}. CARA holds the forecast and host count fixed while repairing resident composition.

Host-group names have no physical meaning. We canonically order groups by their smallest workload identity, then match them to incumbent hosts. Let $n_{hk}$ count residents shared by incumbent host $h$ and candidate group $k$. The minimum number of physical migrations is
\begin{equation}\label{eq:movement}
m(\pi,\inc)=N-\max_{\sigma\in S_H}\sum_{h=1}^{H}n_{h,\sigma(h)}.
\end{equation}
A feasible repair satisfies $m(\pi,\inc)\le K$. The matching also determines the actual destinations and moved identities. Equal objectives use the reference matcher's deterministic tie rule, so an exact query answer includes a physical action as well as a migration count. All experiments use unit movement costs, the information available in the benchmark.

\subsection{Reliability Profiles and Ranked Results}
CARA retains CloudCons' violation rate (VR), violation severity (VS), and peak-to-average ratio (PAR)~\cite{zhang2026cloudcons}. Define $o_{bht}=\mathbf1\{\ell_{bht}>C\}$, $e_{bht}=[\ell_{bht}/C-1]_+$, active-host count $A_b=\sum_h\mathbf1\{\max_t\ell_{bht}>0\}$, violating-time count $n_{bh}=\sum_to_{bht}$, affected-host count $M_b=\sum_h\mathbf1\{n_{bh}>0\}$, and host peak $p_{bh}=\max_t\ell_{bht}$. Then
\begin{align}
\mathrm{VR}_b&=\frac{\sum_{h,t}o_{bht}}{A_bT},&u_b&=\frac{\sum_{h,t}e_{bht}}{A_bT},\label{eq:native1}\\
\mathrm{VS}_b&=\frac1{M_b}\sum_{h:n_{bh}>0}\frac{\sum_te_{bht}}{n_{bh}},\label{eq:native2}\\
\mathrm{PAR}_b&=\frac{\max_hp_{bh}}{A_b^{-1}\sum_hp_{bh}}.\label{eq:native3}
\end{align}
VS is zero when no host overloads. An all-idle scene has $\mathrm{VR}=\mathrm{PAR}=-1$ and $u=\mathrm{VS}=0$, following the benchmark. Uncapped excess $u$ measures burden over the whole scene, whereas VS conditions on violations. Two further coordinates are scene overload $J_b=\mathbf1\{\exists h,t:\ell_{bht}>C+\tau\}$ and $V_b=H^{-1}\sum_h\mathbf1\{\exists t:\ell_{bht}>C+\tau\}$, the fraction of hosts that cross the same threshold. The six-coordinate profile is $L_b=(J_b,V_b,\mathrm{VR}_b,u_b,\mathrm{VS}_b,\mathrm{PAR}_b)$.

The Build table contains $B$ length-$T$ demand windows and ranks candidates; separate Selection and Audit tables freeze and admit one repair, respectively. With scale $s_e=\max\{|\bar L_e(\inc)|,10^{-8}\}$, let $g_e(a)=\overline{L_e(\inc)-L_e(a)}/s_e$. CARA maximizes
\begin{equation}\label{eq:searchkey}
\operatorname{key}(a)=\left(\operatorname{sort}_{\uparrow}\{Q(g_e(a))\}_{e=1}^{6},-m(a,\inc)\right),
\end{equation}
lexicographically, where $Q(z)=\lfloor2^{40}z\rfloor$ and an ascending canonical assignment breaks ties. This leximin order compares the weakest normalized improvement first~\cite{ogryczak1997lexicographic,bouveret2009leximin,przybylski2017multiobjective,bauss2024improvements}. The pool retains the best $P$ distinct candidates. Quantization stabilizes the ordering; admission uses the unquantized losses.

A seed is a candidate partition used to generate queries, while all movement remains relative to $\inc$. A local family redistributes a mutable set $\mathcal U$ between two groups of that seed with fixed backgrounds $B_0,B_1$:
\begin{equation}\label{eq:family}
I_0(A)=B_0\cup A,\qquad I_1(A)=B_1\cup(\mathcal U\setminus A),\quad A\subseteq\mathcal U.
\end{equation}
Other groups remain unchanged. Empty groups are excluded. When both backgrounds are empty, fixing the smallest mutable identity to the first side removes complementary label symmetry.

Let $\mathcal F_{\rm cap}$ contain the nonempty, forecast-feasible canonical assignments in a family. For input pool $\mathcal R$ and previously seen identities $\mathcal S$, the query returns
\begin{align}\label{eq:familyquery}
\mathcal E&=\{a\in\mathcal F_{\rm cap}\setminus\mathcal S:m(a,\inc)\le K\},\nonumber\\
\mathcal R'&=\operatorname{Top}_P(\mathcal R\cup\{r(a):a\in\mathcal E\}),\nonumber\\
\mathcal S'&=\mathcal S\cup\mathcal F_{\rm cap}.
\end{align}
Each record $r(a)$ contains its assignment, profile means, gains, key, and physical mapping. Preserving $\mathcal S'$ matters because later families can depend on the current answer. Selection evaluates $\mathcal R'$ on a separate table, requires coordinatewise noninferiority and one positive gain, and chooses by coordinate midranks, movement, and deterministic ties. It freezes one action for admission. Failed selection or admission leaves $\inc$ in place.

\subsection{Numerical Semantics}\label{sec:numerics}
Capacity tests and rank ties make numerical equivalence part of query semantics. Reassociating a sum can change a predicate at capacity even when the numerical discrepancy is small~\cite{muller2018reproducible}. CARA uses the same ordered binary64 resident, time, and host reductions as its reference evaluator. Shared views reproduce those reductions; pruning uses outward bounds on their stored results. We assume finite nonnegative inputs, finite canonical intermediates, nearest-even arithmetic, at most $2^{50}$ workloads, and exactly represented count normalizers. The derivations in the code artifact specify the reduction trees and rounding enclosures. This contract permits different physical plans to return exactly the same ordered records.

\section{Compiling Related Repair Queries}\label{sec:family}
We organize movement eligibility around reusable keys and seed constraints, then recover physical actions for retained results. Theorem~\ref{prop:core} bounds the remaining assignment core.

\subsection{Execution Overview}\label{sec:overview}
Figure~\ref{fig:method}(a) follows Build queries to one released repair. Search keeps the incumbent fixed and updates a ranked pool and seen relation through virtual seeds. Selection freezes one physical action on its separate table; Section~\ref{sec:admission} admits it or retains the incumbent.

Figure~\ref{fig:method}(b) expands the compiler. Resident identities share load views, origin counts share movement queries, and seed certificates fix reusable matches. A family restores edited columns and supplies numerical budgets to Section~\ref{sec:selection}. There, optimistic keys select candidate profiles against the exact pool cutoff $\theta$; each acquired profile uses all Build windows. We next develop the reuse keys, matching compiler and physical recovery.

\subsection{Resident Views and Origin-Count Classes}
A family stores its seed, two backgrounds and an ordered mutable set of width $w$. Resident tuples are interned by exact equality and key canonical load views. A candidate uses $H$ group identifiers until pool entry requires its full $N$-label assignment. Tiled assembly combines shared projections in the prescribed time and host reduction order.

Forecast subset sums screen masks before resident tuples are constructed. Outward radii enclose their discrepancy from the canonical resident reduction~\cite{higham2002accuracy}. Only a lower load exceeding stored $C+\tau$ rejects a mask; survivors receive the exact canonical capacity predicate. This screen preserves forecast eligibility while avoiding group construction.

Migration uses a coarser key. Let $n_h$ count mutable residents originating on incumbent host $h$, $a_h(A)$ count those sent to the first edited group, and $c_{h0},c_{h1}$ denote background overlaps. The edited overlap columns are
\begin{equation}\label{eq:countsignature}
W_{h0}(A)=c_{h0}+a_h(A),\qquad W_{h1}(A)=c_{h1}+n_h-a_h(A).
\end{equation}
All other columns remain fixed in seed order. The count vector $\boldsymbol a(A)$ determines the movement objective, with at most
\begin{equation}\label{eq:signaturecount}
S_{\mathcal U}\le\prod_{h:n_h>0}(n_h+1)
\end{equation}
distinct queries for $2^w$ masks. Canonical column permutations leave this objective unchanged. CARA solves each count class once, retaining separate workload views for its members. Symmetry-fixed residents are absorbed into the backgrounds before the free counts are formed.

\subsection{Seed-Universal Matching Compilation}
Let $W^0$ be the overlap matrix between the incumbent and a seed $\pi^0$, $s_i$ its incumbent row sum, and $t_j$ its seed column sum. Write $W^a$ for a family member's overlap matrix. For a family leaving column $j$ unchanged, every alternative in row $i$ is at most $s_i-W^0_{ij}$ and every alternative in column $j$ is at most $t_j-W^0_{ij}$. Hence a positive seed edge $z=W^0_{ij}$ is universally fixable when
\begin{equation}\label{eq:seed-edge}
3z\ge s_i+t_j.
\end{equation}
Indeed, if an optimum matches $(i,k)$ and $(\ell,j)$ instead, replacing them with $(i,j)$ and $(\ell,k)$ changes its weight by at least $3z-s_i-t_j\ge0$. The certificate depends only on the seed and incumbent, not on the mutable set or the contents of other edited columns.

CARA scans the sparse seed overlaps once and selects a maximal disjoint set $F$ of such edges. Its positive certificate graph is a union of stars, so this matching is also maximum-cardinality. A family editing columns $J$ restores only edges incident to $J$, giving $F_J=\{(i,j)\in F:j\notin J\}$. At most $p=|J|$ edges are restored. The family optimum is
\begin{equation}\label{eq:factor}
v(a)=\sum_{(i,j)\in F_J}W^0_{ij}+v\bigl(W^a[R_J,C_J]\bigr),
\end{equation}
where $R_J,C_J$ are the remaining rows and columns. Fixed edges and their contribution are shared across families; specialization does not copy the complete certificate list.

\begin{theorem}[Migration-dependent core]\label{prop:core}
For unit movement costs, suppose each incumbent host has at least $q\ge1$ residents and the seed's minimum migration count is $m_0=m(\pi^0,\inc)$. A family editing $p$ columns has residual square dimension
\begin{equation}\label{eq:family-core-budget}
k\le\min\left\{H,\ p+\left\lfloor\frac{3m_0}{q+1}\right\rfloor\right\}.
\end{equation}
\end{theorem}
\noindent\emph{Proof.} Align the seed optimally as a witness. On aligned edge $i$, retained, outgoing, and incoming counts satisfy $s_i=d_i+a_i$, $t_{\sigma(i)}=d_i+b_i$, and $\sum_i a_i=\sum_i b_i=m_0$. If $d_i\ge a_i+b_i$, then $3d_i\ge s_i+t_{\sigma(i)}$, certifying that edge. For the remaining rows $B$, integer counts imply $d_i+1\le a_i+b_i$ and therefore
\[
(q+1)|B|\le\sum_{i\in B}(s_i+1)\le\sum_{i\in B}(2a_i+b_i)\le3m_0.
\]
The certified aligned edges form a matching of size $H-|B|$, so $|F|\ge H-|B|$. Restoring at most $p$ edges leaves $k\le H-|F|+p\le|B|+p$.\hfill$\square$

The compiler does not compute the witness alignment. Construction scans $N$ residents and $E$ positive seed overlaps in $O(N+H+E)$ time and storage. For a family, restoring endpoints and building the residual envelope costs $O(k^2+wp)$ before local reductions and assignment. Identity enumeration, workload evaluation, and physical output construction remain separate costs.

\subsection{Local Reduction, Replay, and Physical Outputs}
Seed certificates are deliberately independent of the edited pair. A tighter envelope $W^+$ on the residual uses its backgrounds and mutable origins. On unchanged edge $(i,j)$, the family fixes it if
\begin{equation}\label{eq:family-core-edge}
W^0_{ij}\ge\max_{k\ne j}W^+_{ik}+\max_{\ell\ne i}W^+_{\ell j}.
\end{equation}
The same exchange argument applies simultaneously to every family member~\cite{grossmann2026hypergraph}. Disjoint certified endpoints are removed in passes. The remaining positive support is decomposed into components; unchanged components contribute constants, while components containing an edited column vary by count class. Rectangular components include zero entries, with surplus endpoints joined by zero edges.

Small residuals use direct assignment~\cite{kuhn1955hungarian,jonker1987assignment,crouse2016assignment}. Larger residuals reuse a seed optimum and feasible duals: changing at most two columns removes their matches and requires at most two shortest augmenting paths to restore optimality~\cite{millstettey2007dynamic,morita2024update}. For padded component dimension $d$, replay costs $O(d^2)$ and $O(d)$ extra working memory.

For retained records, CARA recovers physical destinations from the compiled matching only if all fixed edges are strict and the residual optimum is unique. Cached seed alternative maxima sharpen this check on unchanged endpoints; mutable rows use conservative bounds. Otherwise the original full matcher recovers its deterministic physical tie. Count-class sharing therefore preserves distinct physical records.

\section{Shared Certificates for Profile Acquisition}\label{sec:selection}
Section~\ref{sec:family} establishes movement eligibility. We now select which eligible candidates need exact Build-table profiles. Host summaries and subset sums supply reusable operands; family-wide error budgets turn them into profile bounds; progressive acquisition uses those bounds to preserve the ranked result.

\subsection{Incremental Common-Host Summaries}
Canonical group projections include the per-scene statistics for all six losses. Four additive counts and four numerical summaries cover unchanged hosts. A stable-leaf aggregate tree stores exact integer counts and outward numerical sum intervals. Internal intervals enclose the real sum of their children; native canonical reductions remain the reference for exact profiles.

For an edited pair, the root minus the two leaf contributions encloses the unchanged-host totals in $O(B)$ numerical work. Each node also stores its three largest host peaks with distinct real leaf identifiers. Excluding two leaves leaves an exact unchanged maximum among those three, including ties. Background enclosures use the two edited load projections and the mutable residents' demand rows. Subtracting lower bounds from upper totals, and conversely, preserves the enclosure; the canonical reduction's rounding radius is applied when assembling a metric bound.

On a new seed, unchanged resident tuples retain their slots. With $r$ changed groups and fanout $f$, updating ancestor nodes costs $O(rBf\log_fH)$ after obtaining changed canonical projections. Membership identification can cost $O(N)$; cache misses additionally require resident-load and time projections. Each family retains detached summaries, so subsequent seed changes preserve its certificates.

\subsection{Sharing Mutable Subset Sums}
A profile bound also needs the demand assigned to each edited host. Computing these sums independently for every mask repeats the same additions. CARA uses a bounded meet-in-the-middle representation~\cite{horowitz1974partitions}. Let $b=w-\mathbf1\{\text{a resident is symmetry-fixed}\}$ be the number of free bits, and split those residents into halves $U_0,U_1$. For half $U_j$, CARA builds
\begin{equation}\label{eq:half-subsets}
S_j[0]=0,\qquad S_j[z]=\operatorname{RN}\bigl(S_j[z\setminus\{i\}]+X_i\bigr),
\end{equation}
where $i$ is a deterministic set bit. Each mask then uses two table reads and one addition per scene-time cell. The tables have $2^{\lfloor b/2\rfloor}+2^{\lceil b/2\rceil}$ rows, rather than $2^b$. A symmetry-fixed resident is added once to the first side.

Each subset uses at most $w-1$ inexact nonzero additions, giving a common positive-reduction error bound~\cite{higham2002accuracy}. Tables supply certificate operands; exact profiles use canonical views. Construction is lazy and capped at 32\,MiB; cheap or oversized requests use bounded matrix aggregation under the same enclosure contract.

\subsection{Compiling Load Enclosures Once per Family}
The numerical preparation of a certificate can dominate the exact profiles it saves. A direct enclosure inverts the subset reduction, adds a background interval, and transfers that interval to the canonical resident reduction at every candidate--scene--time cell. CARA compiles the associated error budgets once for the whole family.

For one edited side and scene--time cell, let $\beta_{\rm lo}\le\beta\le \beta_{\rm hi}$ enclose the exact-real background sum, $S$ be the real mutable-subset sum, $s$ its stored answer, and $c$ the canonical stored load. The family compiles a demand upper view $S_{\max}$ covering every mutable subset. Standard nonnegative reduction bounds~\cite{higham2002accuracy} then supply uniform allowances $E_s$ for $|s-S|$, $E_c$ for $|c-(\beta+S)|$, and $E_a$ for the center addition. With upward evaluation during compilation, set
\begin{align}
\rho_-&=\operatorname{Up}(E_s+E_c+E_a),\nonumber\\
\rho_+&=\operatorname{Up}(\rho_-+\beta_{\rm hi}-\beta_{\rm lo}).
\label{eq:compiled-radii}
\end{align}
Each candidate obtains its load endpoints using ordinary nearest-even operations:
\begin{align}
p&=\operatorname{RN}(\beta_{\rm lo}+s),\nonumber\\
L&=\max\{0,\operatorname{RN}(p-\rho_-)\},\quad
U=\operatorname{RN}(p+\rho_+).
\label{eq:compiled-load}
\end{align}
The budgets give $p-\rho_-\le c\le p+\rho_+$. Since $c$ is representable, monotonicity gives $\operatorname{RN}(p-\rho_-)\le\operatorname{RN}(c)=c\le\operatorname{RN}(p+\rho_+)$. Thus $L\le c\le U$ without per-candidate directed-rounding operations. The endpoints target the prescribed stored load; exact profiles continue to use their canonical reductions. Compilation takes $O(wBT)$ work and $O(BT)$ payload; enclosing $q$ leaves still takes $O(qBT)$, besides subset construction and exact profiling. Compilation checks the finite domain of every family member. Unsupported inputs and nonleaf ranges use direct outward enclosures.

\subsection{Coarse and Refined Profile Bounds}
The compiled radii make a coarse six-coordinate profile inexpensive enough to precede the full native calculation. Guaranteed threshold crossings bound $J$, $V$ and VR. Peak loads and active-host counts bound PAR. For excess and VS, the coarse stage avoids constructing every host--time excess value. If $p_h^-$ is a lower stored peak and $n_h^+$ an upper violating-time count, its peak-implied excess $e_h^-$ gives
\begin{equation}\label{eq:peak-excess}
\begin{aligned}
\operatorname{RN}(e_h^-/T)&\le\text{mean excess}_h,\\
\operatorname{RN}\!\left(\frac{e_h^-}{\max\{1,n_h^+\}}\right)&\le\text{severity}_h.
\end{aligned}
\end{equation}
Here $e_h^-$ bounds the stored native transform $[p_h^-/C-1]_+$ downwards. The severity count uses the strict native threshold $C$; tolerant crossings at $C+\tau$ supply a conservative VR numerator. Each host can attain its peak at a different time. The inequalities require no coincidence assumption.

A family also compiles a common lower bound from conservation across its edited pair:
\begin{align}\label{eq:coupled}
[d_0/C-1]_++[d_1/C-1]_+&\ge[(d_0+d_1)/C-2]_+,\nonumber\\
\max(d_0,d_1)&\ge(d_0+d_1)/2.
\end{align}
Combining it with the peak bounds strengthens coarse excess and severity. The same unchanged-host summaries and load radii serve both stages. Both stages bound the same six losses. Refinement computes candidate-specific time-level excess and violation counts; exact profiling follows only when the enclosures cannot exclude a candidate. For active-host bounds $A^-,A^+$, affected-host upper bound $M^+$, violating-time upper bounds $n_h^+$, lower excess sums $E_h^-$, overload-cell count $O^-$, maximum peak $P^-$, and peak-sum upper bound $S_p^+$, representative real-arithmetic bounds are
\begin{align}\label{eq:metricbounds}
\mathrm{VR}^-&=\frac{O^-}{A^+T},&u^-&=\frac{\sum_hE_h^-}{A^+T},\nonumber\\
\mathrm{VS}^-&=\frac{\sum_hE_h^-/\max\{1,n_h^+\}}{\max\{1,M^+\}},&\mathrm{PAR}^-&=\frac{A^-P^-}{S_p^+}.
\end{align}
Rounding allowances transfer these inequalities to the stored native projections. Universal coordinate bounds cover unavailable coarse refinements, including the idle sentinel; an unavailable refined enclosure requests exact scoring. The accompanying derivations give every rounding and denominator step.

Either stage supplies a lower stored profile $L^-_{be}$ and an optimistic key
\begin{equation}\label{eq:optimistic}
U(a)=\left(\operatorname{sort}_{\uparrow}\{Q(g_e^+(a))\}_{e=1}^{6},-m^-(a)\right),
\end{equation}
where $g_e^+=\operatorname{Gain}_e(L(\inc),L^-)$ and $m^-$ is exact for a count-class leaf. Coordinatewise monotonicity of gains, sorting, and quantization gives $\operatorname{key}(a)\le U(a)$. Both stages reject only when $U(a)<\theta$, where $\theta$ is the worst key in a full exact pool; equal keys reach the assignment tie-break.

\subsection{Progressive Acquisition and Continuation}
Small eligible sets use refined-key scheduling, preceded by J/V screening when a cutoff is available. Large sets can establish an exact cutoff with a pilot before coarse-to-refined acquisition; the compiled load radii serve both plans. This choice depends on candidate count, independently of the assignment-core dimension.

For a large family with an incomplete input pool, CARA first evaluates a pilot of at most $P$ eligible candidates in movement order through the ordinary exact consumer. Only a full pool of distinct records establishes a cutoff. Pilot candidates are removed from subsequent requests and never rescored. If the input pool is already full, coarse acquisition starts from its existing cutoff.

The coarse stage evaluates a deterministic sample of 64 remaining masks. When fewer than one eighth are strict losers, the query can bypass further coarse acquisition and prepare refined keys directly. Every sampled loser still has its own valid certificate; the sample supplies no inference about an unchecked mask. Otherwise the coarse stage screens the remaining masks, and only survivors acquire refined optimistic keys. These keys globally schedule exact batches; each retained batch can strengthen the cutoff. Acquisition stops when every remaining upper key is strictly below it. A missing bound remains eligible for exact evaluation. No predicted score establishes the cutoff~\cite{fagin2001aggregation,sui2026cas}.

\begin{theorem}[Exact compiled queries]\label{thm:query-exact}
Under the numerical contract in Section~\ref{sec:numerics}, accepted compiled leaf bounds satisfy $L\le c\le U$. For a common valid input pool and seen relation, CARA returns exactly $(\mathcal R',\mathcal S')$ in Eq.~\eqref{eq:familyquery}, including canonical profiles and reference physical ties.
\end{theorem}
\noindent\emph{Proof.} Equations~\eqref{eq:compiled-radii}--\eqref{eq:compiled-load} enclose every accepted stored load. Optimistic keys upper-bound canonical keys, so strict exclusions preserve top-$P$ retention. Exact eligibility and physical recovery supply the same records, and every forecast-feasible identity enters $\mathcal S'$, including migration and score rejections.\hfill$\square$

Identity enumeration remains part of query cost. Equivalent outputs preserve continuation for a scheduler driven by the pool and seen relation; a CPU-budgeted scheduler can complete different families because execution costs differ.

CARA chooses the acquisition plan using query shape and sampled certificate selectivity. The structural rule uses $H+Tr_{\rm pair}$, where $r_{\rm pair}$ counts all residents of the edited groups, and requires at least 32 eligible masks; its threshold is selected on independent synthetic development data. Families with more than 1,024 eligible masks can use coarse acquisition. An incomplete pool uses the exact pilot only when $qT\ge P(H+Tr_{\rm pair})$, where $q$ is the eligible mask count; smaller queries retain overload-indicator screening and full-key scheduling. These choices affect execution cost, while every exclusion still follows its own exact movement predicate or optimistic key.

\section{One-Action Admission}\label{sec:admission}
The preceding sections construct a ranked pool; Selection now freezes one physical repair. Historical admission evaluates a finite Audit table, while fresh admission tests population targets with independent evidence. We first give the historical gate, then derive action-dependent supports and the fresh release guarantee.

Historical admission compares the six profile coordinates on the complete Audit table. The total $T_e=\sum_b[L_{be}(\inc)-L_{be}(\widehat\pi)]$ is accumulated exactly as a sum of stored dyadic rationals, avoiding rounded pairwise subtraction. The gate admits when all totals are nonnegative and at least one is positive. Immutable workload/time keys share repeated observations across overlapping windows~\cite{tangwongsan2015reactive,traub2018scotty,tangwongsan2021daba}.

An independent fresh-scene source instead supports population admission after adaptive search~\cite{cox1975datasplitting,thomas2015hcpi,laroche2019spibb,wasserman2020universal,laufergoldshtein2023pareto,angelopoulos2025ltt}. Conditional on all information $\mathcal F_0$ fixing the action, targets, and supports, fresh scenes are iid. Its four losses are $(J,V,R,u_H)$, where $R=(HT)^{-1}\sum_{h,t}\mathbf1\{\ell_{bht}>C+\tau\}$ and $u_H=(HT)^{-1}\sum_{h,t}[\ell_{bht}/C-1]_+$ use the fixed $HT$ host--time cells. If $k$ host memberships change, $V$ and $R$ gains have support $k/H$, up to stored-ratio rounding. Common resident groups keep their old physical hosts, so $k$ is determined by the matched repair.

Excess has a finer structural bound. Orient each migrated workload from its old host to its new host, with a declared population demand bound $b_i$. For each positive-support connected component $D$, let $v_D$ be its host count and $\overline d_D$ an upper bound on its maximum directed incoming cut. The component conserves its total demand. Its excess difference is bounded both by $\overline d_D/C$ and by $v_D-1$, hence the stored gain satisfies
\begin{equation}\label{eq:rawsupport}
|G_{u_H}|\le B_{u_H}=\frac1H\sum_D\min\{\overline d_D/C,v_D-1\}+2e_u,
\end{equation}
where $e_u$ bounds one canonical projection's rounding error. Exact cuts are enumerated for small components; larger ones use conservative envelopes. This component bound is never larger than applying the same cuts with a single global $(H-1)/H$ conservation cap. Unchanged hosts contribute no physical gain.

Let $q_e$ be a frozen target and $B_e$ a valid support for the exact difference $G_e$ of stored losses. Normalize $Y_{ei}=(G_{ei}-q_e)/(B_e+|q_e|)\in[-1,1]$ when the denominator is positive, and set it to zero otherwise. With fixed dyadic bets $0<\lambda_j<1$ and nonnegative weights summing to one, use
\begin{equation}\label{eq:mixture}
E_e(t)=\sum_{j\ge0}w_j\prod_{i=1}^{t}(1+\lambda_jY_{ei}).
\end{equation}
An expanding finite subset of this fixed mixture is evaluated with directed enclosures; new components replay the observed prefix. Coordinate $e$ locks when the lower sum reaches $1/\alpha$. All four locks release the action.

\begin{theorem}[One-action false-release control]\label{thm:release}
Let one physical action, its targets and valid almost-sure gain supports be fixed by $\mathcal F_0$. Suppose fresh scenes are iid conditional on $\mathcal F_0$. If any required target is false, $\E[G_e\mid\mathcal F_0]\le q_e$ for some $e$, then for $0<\alpha<1$,
\begin{equation}\label{eq:falseadmit}
\Prb(\text{false release}\mid\mathcal F_0)\le\alpha.
\end{equation}
Dependence across the four coordinates is unrestricted.
\end{theorem}
\noindent\emph{Proof.} For a false coordinate, the full fixed mixture is a nonnegative supermartingale starting at one, and its computed sum is a pathwise lower bound. Joint release requires this coordinate to cross $1/\alpha$; Ville's inequality bounds that event by $\alpha$~\cite{shafer2011martingales,howard2021confidence,vovk2021evalues,ramdas2023gametheoretic,waudbysmith2024betting}.\hfill$\square$

Shared paired replay computes unchanged resident groups once; action supports control how many fresh scenes the frozen decision acquires. Complete proofs and numerical enclosures are in the code artifact.
\section{Experimental Evaluation}\label{sec:experiments}
We evaluate exact-query cost, completed work and released quality under CPU budgets, then admission evidence. Fixed-query comparisons preserve complete outputs; budgeted comparisons measure the controller's released actions under common constraints.

\subsection{Workloads, Controls, and Measurement}
\noindent\textbf{Data.} We use every workload in the four CloudCons consolidation cohorts and the four hourly sources (Table~\ref{tab:data}), preserving CPU units and native metrics~\cite{zhang2026cloudcons}. The hourly releases have workload identities but no observed physical deployment assignment; query placements are therefore trace-derived what-if workloads. Every hourly query uses all complete eight-hour windows in the common observed interval. Ten missing Huawei observations are interpolated within their series.

\begin{table}[t]
\centering\small
\caption{Complete CloudCons inputs. Decision horizons are all available nonoverlapping target days; query windows cover every hourly-source series over its common interval.}
\label{tab:data}
\begin{tabular}{lrrrr}
\toprule
&\multicolumn{2}{c}{Consolidation}&\multicolumn{2}{c}{Hourly queries}\\
Source&VMs&Target days&Series&Windows\\
\midrule
Azure&150&3&10,800&181\\
Borg-d&127&10&414&182\\
Borg-e&68&10&618&181\\
Huawei&575&3&174&205\\
\bottomrule
\end{tabular}
\end{table}

\noindent\textbf{Decision protocol.} Before origin $o$, Build, Selection, and Audit span $[o-192,o-64)$, $[o-64,o-32)$, and $[o-32,o)$, including every complete eight-hour window: 121, 25, and 25 rows. A common trailing-24-hour mean forecast supplies FFD and the capacity predicate. Capacity is four for Azure and Huawei and one for the Borg cohorts. The next 24 hours form the target. Each repair retains the host count and may move at most 20\% of workloads. All 25 released origins $288,\ldots,312$ and three search streams give 300 primary decisions. A second view covers all 26 nonoverlapping cohort-days from origin 264, giving 78 decisions. Accounting for 24 shared cases leaves 354 unique contexts. Search streams reuse the same observed targets; summaries weight the four cohorts equally and retain every incumbent fallback.

\noindent\textbf{Query plans.} Eager constructs complete candidates with shared canonical host views; Compact adds compact partitions and origin-count matching while acquiring every eligible profile. Sparse-JV uses a global sparse Jonker--Volgenant assignment per count class~\cite{jonker1987assignment}; private dummy columns retain zero-overlap destinations. Component factors only positive overlap support. Static-Reduction applies neighborhood removal separately to each count matrix~\cite{grossmann2026hypergraph}. Dynamic replay updates a global optimum after the two-column edit~\cite{millstettey2007dynamic,morita2024update}. Family-local CARA compiles each family separately, with an acquisition rule based on host count. Gate-only applies the new structural rule to that same executor. The four leave-one-component-out plans disable seed compilation, common-host summary views, the preliminary screens, or half-subset views, respectively. A separate two-by-two control crosses compiled radii and staged acquisition while retaining all four shared representations. Assignment substitutions share CARA's structural policy and numerical acquisition path. Each leave-one-component-out plan changes only its named component; the two-by-two control separates compiled radii from staged acquisition.

\noindent\textbf{Repair policies.} Swap uses moves and swaps; Anneal minimizes uncapped excess~\cite{kirkpatrick1983annealing}; Joint-SA minimizes mean normalized VR, VS, PAR and excess, and Risk-SA combines host-violation frequency with normalized excess. They share geometric cooling and balanced retention. Broad uses destroy-and-repair proposals. Scenario-MIP minimizes mean excess on four evenly spaced Build scenes, with common capacity and migration constraints, a 1\% gap and one HiGHS thread. Eager-, Compact-, Component-, and Family-CARA use the corresponding executor within the same controller. No-residual, No-frontier, Best-seed, and Raw-order remove complete-family refinement, frontier initialization, portfolio scheduling, and balanced retention. Every policy shares the forecast, Selection, physical validation, Audit, and native target evaluator. Baseline settings are fixed across cohorts.

\noindent\textbf{Development.} The controller grid uses origins 216, 228 and 240, whose targets precede evaluation, on all cohorts and two streams. It selects width twelve and a half-budget swap frontier from widths $\{8,10,12\}$, swap/joint frontiers and fractions $\{0.25,0.5\}$. Independent synthetic studies select batch 16 from four values and threshold 192 from eight values over 24 query shapes; complete grids accompany the code. All plans use $P=16$, the 32\,MiB resident-view cache and bounded assembly; score certificates require 32 eligible masks. Half-subset tables have a separate 32\,MiB cap. Integer studies select assignment and physical-recovery thresholds 64 and 16.

\noindent\textbf{Costs and correctness.} Experiments use an Apple M5 CPU, 16\,GB memory, Python, NumPy, SciPy and one numerical thread. Query timings use three serial rotated trials; tables sum per-context medians. CPU includes session setup, seed compilation, views, predicates, exact scoring, retention and physical outputs. Input loading and window construction are separate common costs. Untimed checks compare full ordered records, canonical NumPy profiles, dense objectives per count class, every excluded leaf and physical destinations. Adaptive cases also compare the continuing pool and seen relation. Separate memory probes exclude the common immutable panel.

Budgeted policies receive two process-CPU seconds for historical representation, setup and proposals. A started operation completes, and its overrun is recorded. Complete decision time adds Selection, validation and Audit; forecast/FFD preparation, source loading and target evaluation are separate. Scenario-MIP charges model construction before its remaining wall-limited solve.

\subsection{Complete-Source Query Execution}
The controlled query workload uses a round-robin partition with $H=\lceil N/8\rceil$, capacity 1.05 times its largest first-day mean forecast load, and eight complete width-eight families per source.

\begin{table*}[t]
\centering\small\setlength{\tabcolsep}{4.1pt}
\caption{Complete-source width-eight queries. CPU includes cold preparation and final physical outputs. The last two columns give Compact/CARA pairs; memory is peak traced allocation beyond the common immutable input.}
\label{tab:queries}
\begin{tabular}{lrrrrrrrrr}
\toprule
& &\multicolumn{6}{c}{CPU (s)}&\multicolumn{2}{c}{Compact/CARA}\\
\cmidrule(lr){3-8}\cmidrule(lr){9-10}
Source & $H$ & Eager & Compact & Sparse-JV & Family & Recomputed & CARA & Profiles & MiB\\
\midrule
Azure & 1350 & 19.678 & 3.902 & 0.874 & 0.672 & 0.485 & 0.372 & 2,040/48 & 163.2/134.6\\
Borg-d & 52 & 0.486 & 0.219 & 0.219 & 0.214 & 0.214 & 0.214 & 2,034/2,034 & 82.5/82.5\\
Borg-e & 78 & 0.639 & 0.255 & 0.211 & 0.248 & 0.338 & 0.201 & 2,004/75 & 82.8/55.0\\
Huawei & 22 & 0.305 & 0.172 & 0.176 & 0.170 & 0.170 & 0.170 & 1,956/1,956 & 85.1/85.1\\
\bottomrule
\end{tabular}
\end{table*}

Table~\ref{tab:queries} holds all 32 families and their rich outputs fixed. Aggregate CPU is 0.957\,s for CARA, 4.549\,s for Compact, 1.480\,s for Sparse-JV and 1.304\,s for Family-local: speedups of 4.75, 1.55 and 1.36, respectively. Azure uses 48 exact profiles instead of 2,040; resident-reduction operands, including the common initial projection, fall from 63.04 to 16.75 million. At an identical 75-profile count on Borg-e, compiling load radii reduces CPU from 0.338 to 0.201\,s; Family-local needs 0.248\,s and 2,004 profiles. Thus the compiler also lowers the cost of obtaining certificates.

The Azure seed index is constructed once and reused seven times. Specializing its eight families reads 16 positive seed-overlap entries; the count queries use 768 local matrix cells on two-row cores. The summary tree occupies 29.77\,MB; total peak traced allocation is 134.6\,MiB, compared with 163.2 for Compact and 530.1 for Eager.

\noindent\textbf{Connected and changing seeds.} Nineteen complete-source settings cover cycle spans 8, 32, 128 or all Azure hosts, full-source cycles elsewhere, equal-migration concentrated/dispersed seeds, and 16-query adaptive sequences per source. All 184 complete answers agree, and every migration-core bound holds. Aggregate CPU is 8.687\,s for CARA, 12.987 for Component, 12.013 for Sparse-JV and 10.325 for Dynamic replay. The full Azure cycle connects all 1,350 hosts, yet leaves two-row cores: CARA takes 0.456\,s, versus 1.632 for Component and 0.719 for Dynamic replay. At equal seed migration count 2,151, dispersed and concentrated Azure seeds leave two and 268--269 core rows, respectively. Connectivity alone therefore does not determine the residual assignment work.

\noindent\textbf{Complete controller paths.} We replay the same completed Eager-CARA operations at origins 288, 300 and 312 on every cohort, holding random draws, annealing progress and cost feedback fixed. These 12 paths contain 170 complete families; Selection and Audit remain inside the decision timer. Every complete answer and admission field agrees. CARA takes 6.765\,s overall, versus 11.500 for Eager, 8.580 for Compact, 6.864 for Sparse-JV, 6.789 for Component and 6.915 for Family-local.

\subsection{Heterogeneous Packings and Wider Families}
A second workload derives feasible FFD placements from the first-day forecast of every hourly-source VM. One common capacity rule is
\begin{equation}\label{eq:packed-capacity}
C=1.05\max\left\{\max_i\widehat x_i,\;\frac8N\sum_i\widehat x_i\right\}.
\end{equation}
The resulting host counts are 489, 50, 48, and 21; maximum occupancies are 310, 40, 226, and 30. Each source has the packed seed and a seed obtained after eight completed adaptive queries on its first 32 windows. Those prefix windows construct the query workload; held-out quality is evaluated separately. Each requested width, eight, twelve and sixteen, uses four pairs spanning small--large, quartile, central and large--large occupancies. A pair uses all $w$ mutable residents whenever available, with unbalanced side counts allowed; smaller complete unions retain their actual width. All measurements use every common-interval window.

\begin{figure*}[t]
\centering\includegraphics[width=\textwidth]{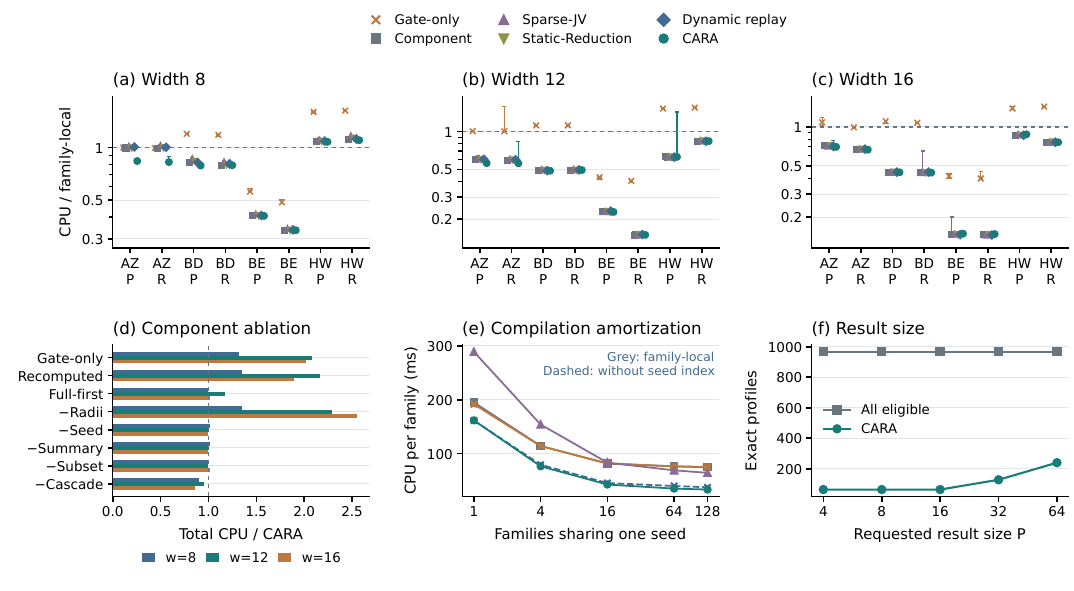}
\caption{Exact query execution with complete source data. (a--c) Packed and repair-derived seeds at requested widths 8, 12 and 16; each source appears in both seed forms. Assignment substitutions share CARA's numerical acquisition path. CPU is relative to the family-local executor; whiskers span three serial rotations. (d) The radii/staging cross and shared-view removals use the same structural policy. (e) Cold and amortized execution on strict prefixes of one fixed 128-family connected-seed sequence. (f) Exact profiles acquired as requested result size $P$ varies on an independent synthetic workload. Every plotted plan returns the same complete ranked physical answer for its query workload. P/R denote packed/repair-derived seeds; AZ, BD, BE and HW identify the four sources in Table~\ref{tab:data}.}
\label{fig:execution}
\end{figure*}

All 96 families return identical complete records under all 14 plans. At requested widths eight, twelve and sixteen, speedups over Family-local are 1.53, 3.00 and 3.07. Total CPU falls from 117.57 to 38.59\,s, a 3.05-fold speedup. At width sixteen, exact profiles fall from 294,461 to 365; 20 of these 32 families use all sixteen mutable residents, and every shorter family remains in the totals. Sparse, component, static-reduction and dynamic-replay substitutions take 38.49--38.69\,s overall. Their shared numerical path supplies the main packed-query gain; matching reuse becomes more consequential on connected seeds and repeated families.

\noindent\textbf{Component attribution.} Gate-only takes 77.63\,s, separating the structural policy from the compiled representation. In the numerical cross, compiling radii reduces full-first CPU from 73.39 to 39.57\,s at the same 683-profile count. Adding staged acquisition reduces it to 38.59\,s; at width twelve the reduction is from 2.759 to 2.344\,s. On the two Borg-e width-sixteen workloads, staging lowers full-first CPU from 7.54/6.37 to 5.63/5.58\,s. Direct refined-key acquisition without either preliminary screen takes 33.52\,s overall: preliminary acquisition pays only when it avoids enough subsequent work. Seed and summary removals are nearly neutral on these short packed-seed sequences; the next study measures their reuse. Disabling half-subset views takes 39.11\,s.

\subsection{Reuse, Acquisition, and Parameters}
\noindent\textbf{Compilation amortization.} The full Azure cycle connects all 1,350 hosts. One fixed sequence of 128 distinct, uniformly spread width-eight pairs defines strict prefixes $1,4,16,64,128$; each starts fresh and charges compilation once. Independent checks cover the complete sequence. At 128 families, CARA takes 4.242\,s, versus 9.539 for Family-local and 8.224 for Sparse-JV. Removing the seed index raises CPU to 4.779\,s; removing summary views raises it to 7.396\,s. The index scans 2,700 seed entries once and specializes the 128 families by reading 512 entries in total, with two-row cores. Cold single-family CPU is 0.162\,s, versus 0.195 for Family-local. The summary tree occupies 29.77\,MB; peak traced allocation at 128 families is 518.5\,MB, versus 507.1 for Family-local, including the returned prefix records.

\noindent\textbf{Result size and acquisition.} An independent workload has 512 hosts, 96 windows and five complete families. Varying $P\in\{4,8,16,32,64\}$ keeps the compiler and batch size fixed. CARA acquires 64, 64, 64, 128 and 240 exact profiles, versus 967 for complete acquisition at every budget. The fixed batch limit explains the flat region through $P=16$. Complete CPU speedups range from 1.85 to 3.12; at $P=16$, CPU falls from 0.378 to 0.129\,s.

\noindent\textbf{Cost policy and controller parameters.} On the independent 24-context grid with recomputed load enclosures, threshold 192 takes 0.309\,s in aggregate median CPU, compared with 0.354\,s for the host-only gate, 0.317\,s for always preparing score certificates and 0.415\,s for never preparing them. It is the minimum in all three rotations; threshold 256 takes 0.313\,s. The selected batch of 16 balances cutoff updates against profile batching.

\subsection{Released Repair Quality}
We report every native CloudCons outcome: hosts, utilization, VR, VS, PAR, PICP, MPIW and Winkler score, together with excess and released movement. Prediction intervals use the benchmark's 95\% construction, 100 host simulations and preceding-24-hour demand scales. Forecast accuracy uses native MASE and normalized nine-quantile CRPS. Target outcomes are evaluated after the actions and admission decisions are fixed.

\begin{table*}[t]
\centering\footnotesize\setlength{\tabcolsep}{3.8pt}
\caption{Primary comparison over all 300 decisions, including incumbent fallbacks. Cohorts receive equal weight. Lower VR, VS, PAR, excess, and Winkler scores are better; minima of these five loss columns are bold. Coverage and interval width are reported jointly. Movement is the percentage of workloads actually moved after admission.}
\label{tab:main}
\begin{tabular}{lrrrrrrrrrr}
\toprule
Method & Hosts & Util. & VR & VS & PAR & Excess & PICP & MPIW & Winkler & Move (\%)\\
\midrule
FFD & 14.67 & 0.94 & 0.4211 & 0.09951 & 1.182 & 0.04463 & 0.7408 & 0.3698 & 1.4241 & 0.00\\
Swap & 14.67 & 0.94 & 0.3065 & 0.08825 & 1.147 & 0.03636 & 0.7485 & 0.3745 & 1.3513 & 11.16\\
Anneal & 14.67 & 0.94 & 0.2846 & 0.08073 & 1.129 & 0.02990 & 0.7496 & 0.3742 & 1.3548 & 14.71\\
Risk-SA & 14.67 & 0.94 & 0.3055 & 0.08424 & 1.136 & 0.03025 & 0.7506 & 0.3750 & 1.3673 & 12.54\\
Joint-SA & 14.67 & 0.94 & 0.2862 & 0.08152 & 1.129 & 0.03010 & 0.7507 & 0.3758 & 1.3504 & 14.55\\
Broad & 14.67 & 0.94 & 0.3063 & 0.08375 & 1.125 & 0.03007 & 0.7512 & 0.3776 & 1.3639 & 14.55\\
Scenario-MIP & 14.67 & 0.94 & 0.3821 & 0.09694 & 1.173 & 0.04165 & 0.7469 & 0.3720 & \textbf{1.3408} & 4.96\\
Eager-CARA & 14.67 & 0.94 & 0.2859 & 0.08145 & 1.130 & 0.02943 & 0.7554 & 0.3771 & 1.3528 & 13.15\\
Compact-CARA & 14.67 & 0.94 & 0.2875 & 0.08334 & 1.128 & 0.02988 & 0.7506 & 0.3764 & 1.3622 & 12.88\\
Component-CARA & 14.67 & 0.94 & 0.2830 & 0.08206 & \textbf{1.124} & 0.02874 & 0.7496 & 0.3755 & 1.3827 & 13.67\\
No-residual & 14.67 & 0.94 & 0.2811 & \textbf{0.07778} & 1.135 & 0.02935 & 0.7511 & 0.3753 & 1.3600 & 13.96\\
No-frontier & 14.67 & 0.94 & 0.3102 & 0.08409 & 1.125 & 0.02952 & 0.7513 & 0.3781 & 1.3710 & 14.69\\
Best-seed & 14.67 & 0.94 & \textbf{0.2807} & 0.08355 & 1.127 & 0.02925 & 0.7507 & 0.3767 & 1.3665 & 14.10\\
Raw-order & 14.67 & 0.94 & 0.3092 & 0.08805 & 1.146 & 0.03617 & 0.7489 & 0.3739 & 1.3451 & 11.22\\
Family-CARA & 14.67 & 0.94 & 0.2820 & 0.08177 & 1.124 & 0.02886 & 0.7510 & 0.3763 & 1.3724 & 13.83\\
\textbf{CARA} & 14.67 & 0.94 & 0.2824 & 0.08181 & 1.125 & \textbf{0.02863} & 0.7496 & 0.3758 & 1.3784 & 13.83\\
\bottomrule
\end{tabular}
\end{table*}

Table~\ref{tab:main} reports all primary decisions. CARA reduces cohort-macro VR, VS, PAR and excess over FFD by 32.9\%, 17.8\%, 4.9\% and 35.8\%, respectively. Its excess is the lowest among the compared policies: 4.9\% below Joint-SA with 5.0\% less movement, and 0.8\% below Family-CARA. CARA combines the lowest excess with lower VR and PAR than Joint-SA. PICP rises from 0.7408 for FFD to 0.7496, with MPIW 0.3758 and Winkler score 1.3784. Every policy retains the same host count and utilization.

Removing complete-family refinement raises excess by 2.5\%; removing the frontier raises it by 3.1\%. Best-seed and Raw-order increase it by 2.1\% and 26.3\%, respectively. Complete decision CPU averages 2.033\,s for CARA, 2.041\,s for Family-CARA, 2.046\,s for Compact-CARA and 2.074\,s for Eager-CARA. CARA scores 389,434 candidates across the primary view, versus 1,070,511 for Family-CARA; the latter completes a similar number of families on these smaller host sets. The separate daily view lowers excess by 19.3\% over FFD and uses 18.6\% less movement than Joint-SA. All 78 daily decisions, including shared primary cases and incumbent fallbacks, contribute to that view.

\noindent\textbf{Full-source budgeted repair.} To connect wider query execution to decisions on all hourly-source VMs, an additional shorter-history study uses every nonoverlapping 24-hour target after hour 96: 13 source-days, or 312 cohort-hours. Build, Selection and Audit have 41, 17 and 17 complete windows in the disjoint intervals $[o-96,o-48)$, $[o-48,o-24)$ and $[o-24,o)$. Capacity follows Eq.~\eqref{eq:packed-capacity}. Swap, Joint-SA, Family-local CARA and CARA receive two- and ten-second CPU budgets, with three streams. A history-only data object ends at the origin, and all methods, budgets and streams are frozen before its target opens. The study adds 78 budget--stream contexts to the long-history comparison.

\begin{figure*}[t]
\centering\includegraphics[width=\textwidth]{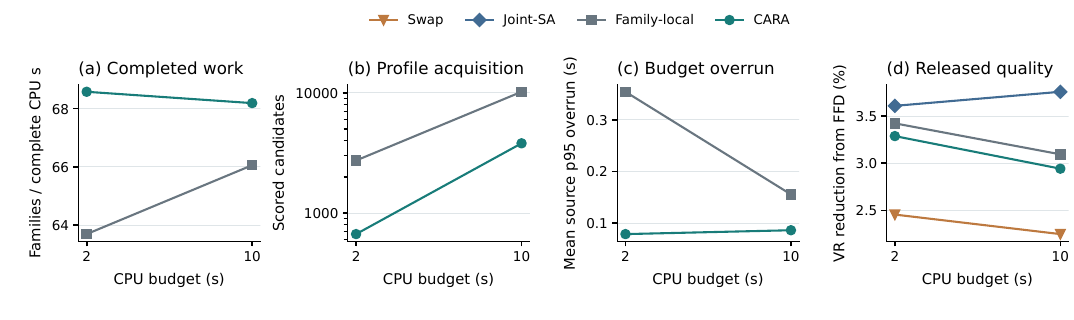}
\caption{Budgeted repair on every hourly-source VM. Panels (a--c) compare the two complete-family executors; (d) also includes Swap and Joint-SA. Each point includes all 13 nonoverlapping source-days and all three streams, with equal source weight and every incumbent fallback. The p95 overrun is averaged across sources; complete CPU includes Selection and Audit.}
\label{fig:quality}
\end{figure*}

CARA completes 142.0 and 687.7 families at the two budgets, compared with 136.4 and 667.3 for Family-local, while scoring 669.7 and 3,802.1 candidates instead of 2,736.2 and 10,185.6. Complete decision CPU is 2.071 and 10.086\,s, versus 2.142 and 10.102\,s. The mean source-level p95 overrun falls from 0.354 to 0.078\,s at the two-second budget, and from 0.155 to 0.086\,s at ten seconds. Released VR improves over FFD by 3.29\% and 2.94\%. Across both budgets, CARA's macro VR, VS, PAR and excess remain within 0.40\% relative of Family-local. Figure~\ref{fig:quality} connects the lower profile-acquisition cost to more completed work and tighter budget adherence.

\noindent\textbf{Common forecast.} Across all 118 distinct long-history cohort--origin targets, MASE is 1.1511, 0.9619, 1.3255 and 1.1820 on Azure, Borg-d, Borg-e and Huawei; normalized CRPS is 0.04028, 0.17054, 0.22815 and 0.05892. Every policy uses these identical forecasts. The 105 Huawei VM--origin pairs with zero lag-24 scale are excluded only from the undefined MASE ratio; all remain in placement metrics and CRPS.

\subsection{Case Study}
\begin{figure*}[t]
\centering\includegraphics[width=\textwidth]{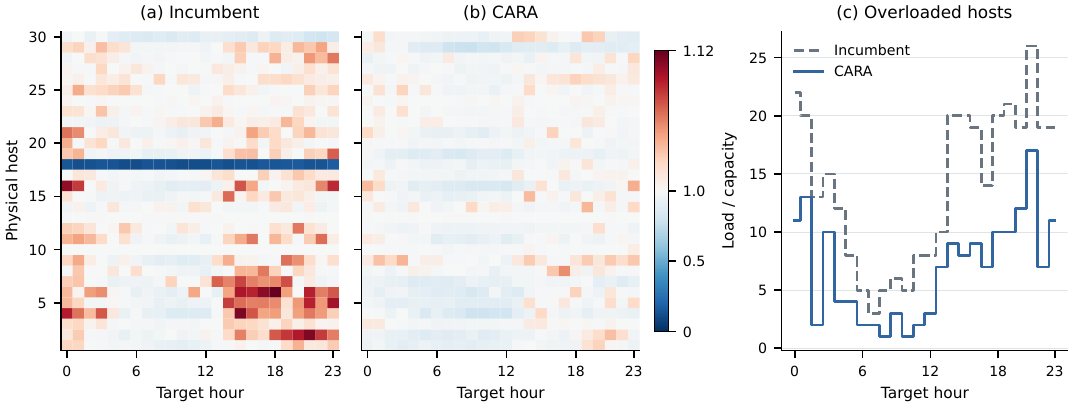}
\caption{Huawei at origin 297, stream zero: all 30 physical hosts and all 24 target hours. Load panels use the same capacity-centered scale and physical ordering. The repair moves 115 workloads within the 20\% limit; overloaded host--hours fall from 337 to 165.}
\label{fig:case}
\end{figure*}

At origin 297, the incumbent uses all 30 hosts at 96.8\% utilization. Its loads coincide near capacity: 337 of 720 host--hours overload, and the busiest target hour overloads 26 hosts. CARA redistributes 115 of 575 workloads while retaining the host budget and utilization. Successive local families can accumulate changes in a seed, with the incumbent still fixing the total migration limit. Overloaded host--hours fall to 165, the busiest hour falls to 17 affected hosts, and excess decreases from 0.01263 to 0.00283. Figure~\ref{fig:case} shows the same physical ordering and capacity-centered scale for both placements. The improvement comes from spreading coincident peaks within the fixed capacity.

\subsection{One-Action Admission}
\noindent\textbf{Historical census.} The census covers all 354 decisions, including 288 selected actions and 66 no-action fallbacks; each selected action uses its complete 25-window Audit table. Direct and source-keyed canonical readers, crossed with Fraction and integer-dyadic accumulation, preserve every admission field and native profile. Preparing the role inputs and reference costs 0.099\,s in total. Additional reader-and-gate CPU sums to 0.104\,s for direct/Fraction execution and 0.060\,s for keyed/dyadic execution. Candidate source-cell fetches fall from 13,572,600 to 2,171,616 because overlapping windows share their workload/time observations. The separate independent oracle is excluded from these execution costs.

\noindent\textbf{Fresh supports and evidence.} The fresh study fixes one action in each of 96 environments: 32 complementary, bursty-common and weakly complementary demand laws each, including six no-action fallbacks. Each has 48 VMs, eight time points, unit capacity and demands clipped to $[0,1]$, giving population bounds $b_i=1$. Shared factors create within-scene dependence; scenes are independently generated. Required gains for $(J,V,R,u_H)$ are $(-0.01,-0.01,0.005,-0.001)$, with $\alpha=.05$, inspection every 32 scenes and a cap of 8,192. All native Evaluation metrics use 16,384 independent scenes per environment.

\begin{table}[t]
\centering\small\setlength{\tabcolsep}{4.5pt}
\caption{Complete fixed-action comparison over 96 environments. Means include six no-action fallbacks. Sequential plans inspect every 32 scenes with the same 8,192-scene cap.}
\label{tab:fresh}
\begin{tabular}{lrrr}
\toprule
Plan & Releases & Mean scenes & CPU (ms)\\
\midrule
Global support & 78 & 1,489.0 & 10.68\\
Without graph bound & 78 & 1,390.7 & 9.94\\
Without Boolean bound & 78 & 1,451.7 & 10.22\\
Without component bound & 78 & 1,347.0 & 9.51\\
Fixed 2,048 & 74 & 1,920.0 & 13.45\\
Fixed 4,096 & 77 & 3,840.0 & 26.15\\
Fixed 8,192 & 78 & 7,680.0 & 51.58\\
\textbf{CARA} & 78 & 1,347.0 & 9.53\\
\bottomrule
\end{tabular}
\end{table}

CARA releases the same 78 actions as Global and Fixed 8,192, using 9.5\% and 82.5\% fewer scenes, respectively; mean CPU is 9.53\,ms, versus 10.68 and 51.58\,ms. Removing the Boolean or graph bound increases mean scenes to 1,451.7 or 1,390.7. The component bound tightens the excess endpoint, while the fixed-host violation rate often finishes last, leaving the overall stopping times tied here. The three 78-release plans have the same terminal native VR, VS and PAR: 0.2690, 0.1445 and 1.1890 on the independent Evaluation population.

Three independent null checks each use 1,024 streams: symmetric $\pm1/2$ gains, rare-positive gains $15/16$ with probability $1/16$ and $-1/16$ otherwise, and one symmetric null with three gains fixed at $1/4$. False releases number 12, 18 and 14; their one-sided 95\% Clopper--Pearson upper bounds are 1.89\%, 2.60\% and 2.13\%.

\section{Related Work}
\noindent\textbf{Workload-aware placement.} Database consolidation and tenant-placement systems use workload observations to balance utilization and reliability~\cite{curino2011kairos,schaffner2013rtp,taft2016step,cortez2017resourcecentral,poppe2020seagull,konig2022tenant,arora2023flexible,konig2023solver}. Chance-constrained packing models uncertain demand~\cite{song2014chance,cohen2019overcommitment,zhang2020branch}; CloudCons evaluates forecasting through downstream consolidation~\cite{zhang2026cloudcons}. CARA addresses repeated incumbent-relative repair queries within this workflow. It shares workload views and movement constraints while retaining the chosen forecast, host budget, and physical migration semantics.

\noindent\textbf{Shared computation and matching.} Multiquery execution, incremental views, and compiled what-if evaluation share related computations~\cite{roy2000multiquery,giannikis2012shareddb,ahmad2012dbtoaster,budiu2023dbsp,kara2024fivm,mohammed2024fade}. Matching reductions and primal-dual updates accelerate individual assignments~\cite{millstettey2007dynamic,korenwein2018reduction,morita2024update,grossmann2026hypergraph}. CARA compiles seed-universal overlap certificates once and specializes them into origin-count families before individual solves. Its migration-dependent core bound and shared numerical certificate views connect matching reuse to the ranked workload consumer.

\noindent\textbf{Selective evaluation.} Adaptive predicates and exact top-$k$ algorithms avoid irrelevant computations~\cite{shewchuk1997adaptive,fagin2001aggregation,ilyas2004rankjoin}. Recent semantic and AI query systems couple selective evaluation with answer guarantees~\cite{lee2026evergreen,wang2026querycentric,hassanzadeh2026semviews,sui2026cas}. CAS certifies scenario-based rankings; SemViews reuses labels to guide fresh semantic-filter verification. CARA combines resident conservation with compiled movement constraints to bound numerical repair profiles. Its acquisition preserves physical outputs and continuation state as well as the ranked result.

\noindent\textbf{Adaptive repair and admission.} Local branching and adaptive neighborhoods generate candidates under limited budgets~\cite{fischetti2003localbranching,ropke2006alns,prudhomme2014explanation,pisinger2019lns,liu2022learninglocal,lewander2026banditlns}. Data splitting and anytime-valid inference support post-selection evaluation~\cite{cox1975datasplitting,howard2021confidence,waudbysmith2024betting,angelopoulos2025ltt}; recent self-evolving agents also freeze updates before fresh acceptance tests~\cite{sengupta2026sea,yang2026selection}. CARA specializes this pattern to physical repairs. Migration components supply action-specific evidence bounds, and the same resident dependencies drive exact query sharing; the proposal policy remains replaceable.

\section{Conclusion}\label{sec:conclusion}
We presented CARA, an exact repair query processor with one-action certification for cloud consolidation. Seed-universal matching and host summaries share work across families, while family-wide numerical budgets lower certificate cost. Direct and staged acquisition preserve ranked physical answers and continuation state. Complete-source packed queries accelerate 3.05-fold; primary decision excess falls 35.8\% over FFD. Physical migration components provide evidence bounds for the release of one frozen repair.

\clearpage
\section*{Acknowledgment}
LLM was used for language polishing. All AI-assisted parts were reviewed and verified by the authors, who take full responsibility for the content of this paper.

\bibliographystyle{IEEEtran}
\bibliography{references}
\end{document}